\def\year{2022}\relax
\documentclass[letterpaper]{article} 
\usepackage{aaai22}  
\usepackage{times}  
\usepackage{helvet}  
\usepackage{courier}  
\usepackage[hyphens]{url}  
\usepackage{graphicx} 
\graphicspath{ {./figs/} }
\usepackage{natbib}  
\usepackage{caption} 
\DeclareCaptionStyle{ruled}{labelfont=normalfont,labelsep=colon,strut=off} 
\usepackage{algorithm}
\usepackage{algorithmic}
\usepackage{subfigure}
\usepackage{amsmath}
\usepackage{multirow}
\usepackage{amssymb}
\usepackage{booktabs}
\usepackage{newfloat}
\usepackage{listings}
\floatstyle{ruled}
\newfloat{listing}{tb}{lst}{}
\floatname{listing}{Listing}
\title{SleepPriorCL: Contrastive Representation Learning with Prior Knowledge-based Positive Mining and Adaptive Temperature for Sleep Staging}
\author{
    Hongjun Zhang,
    Jing Wang,
    Qinfeng Xiao,
    Jiaoxue Deng,
    Youfang Lin
}
\affiliations{
    School of Computer and Information Technology, Beijing Jiaotong University, Beijing, China\\
    Beijing Key Laboratory of Traffic Data Analysis and Mining, Beijing, China\\
}

\usepackage{bibentry}

\begin{document}

\maketitle

\begin{abstract}

%
The objective of this paper is to learn semantic representations for sleep stage classification from raw physiological time series. Although supervised methods have gained remarkable performance, they are limited in clinical situations due to the requirement of fully labeled data. Self-supervised learning (SSL) based on contrasting semantically similar (positive) and dissimilar (negative) pairs of samples have achieved promising success. However, existing SSL methods suffer the problem that many semantically similar positives are still uncovered and even treated as negatives. In this paper, we propose a novel SSL approach named SleepPriorCL to alleviate the above problem. Advances of our approach over existing SSL methods are two-fold: 1) by incorporating prior domain knowledge into the training regime of SSL, more semantically similar positives are discovered without accessing ground-truth labels; 2) via investigating the influence of the temperature in contrastive loss, an adaptive temperature mechanism for each sample according to prior domain knowledge is further proposed, leading to better performance. Extensive experiments demonstrate that our method achieves state-of-the-art performance and consistently outperforms baselines.

%
\end{abstract}

\section{Introduction}

Identifying sleep stages \cite{aboalayon2016sleep} is essential for evaluating sleep quality and diagnosing sleep disorders. Traditionally, sleep staging is finished by well trained experts according to physiological signals, which is laborious and time-consuming. Thus various supervised methods
\cite{Automaticsleepstaging, GraphSleepNet} are developed to automate sleep staging. Those approaches can be categorized into two paradigms: (i) handcrafted feature based machine learning classifiers with strong interpretability; (ii) end-to-end deep neural networks with better performance but worse interpretability. However, both of the two paradigms require fully labelled datasets, which are laborious to acquire in health care area. Recent progress of self-supervised learning (SSL) \cite{jing2019selfsupervised,he2020momentum,chen2020simple} has gained promising performance for physiological time series, with competitive performance compared with supervised methods \cite{franceschi2019unsupervised,DBLP:conf/icassp/XiaoWYZBZW21}. SSL methods can extract representations with semantic information from raw physiological signals, which are promising to alleviate the burden of manual labeling works.
\begin{figure}[t]
\centering
\includegraphics[width=0.9\columnwidth]{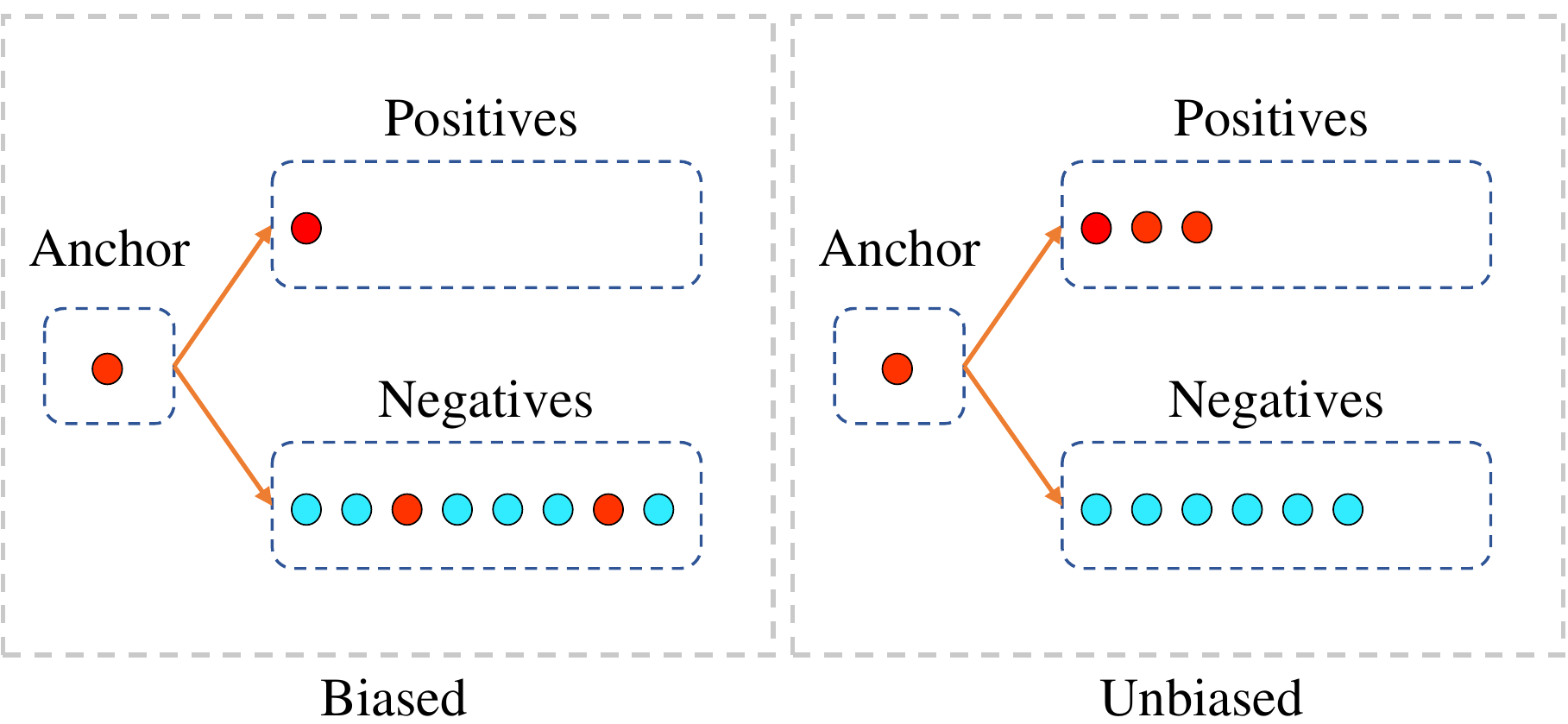} 
\caption{Sampling bias in contrastive learning. Traditional contrastive learning is biased,  which ignores many potential positives. It can be unbiased if given ground-truth labels, but such operation loses the advantage of self-supervised learning. Our goal is to alleviate sampling bias problem without labels.}
\label{biasedsample}
\end{figure}

Many efficient self-supervised learning methods enable neural networks to learn semantically meaningful representations through instance discrimination on data samples, e.g. contrastive learning \cite{chen2020simple, he2020momentum}. In detail, given an anchor sample, semantic-similar positives are attracted in representation space while others (a.k.a. negatives) are repelled. However, negatives are typically sampled randomly, which inevitably contains potential positives. As mentioned in recent literatures \cite{DCL, tonekaboni2021unsupervised}, we refer to this problem as \emph{sampling bias}. 

\begin{figure}[t]
\centering
\includegraphics[width=0.8\columnwidth]{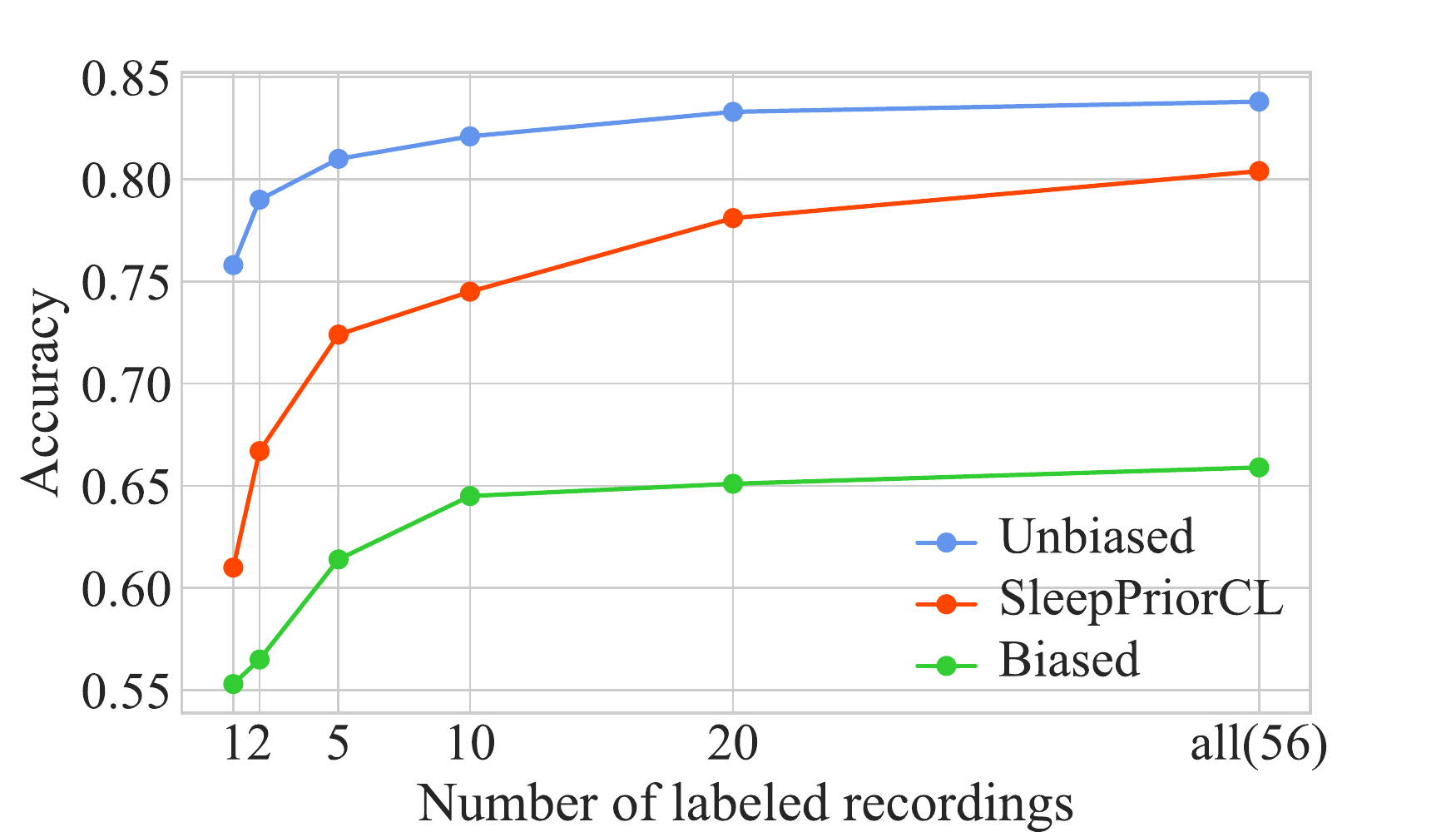}
\caption{SleepPriorCL reduce the performance gap caused by sampling bias. Experiment on MASS-SS3.}
\label{biased}
\end{figure}

Instance-discrimination based approaches suffer the sampling bias problem. As Figure \ref{biasedsample} shows, traditional contrastive learning \cite{chen2020simple} obtains the only positive by semantic-invariant augmentation, which ignores a wider range of other semantically similar positives. What's more, a line of time series self-supervised learning methods  consider sample positives according to local smoothness of time series \cite{10.1088/1741-2552/abca18,tonekaboni2021unsupervised}, i.e. temporal neighbors are considered as positives. However this criterion can also lead to severe bias since semantically similar samples are not necessarily temporally neighboring, e.g., the same sleep stage occurs in different sleep cycles. Sampling bias leads to a performance drop. As showed in Figure \ref{biased}, there is a large gap between unbiased and biased sampling Mechanisms. In this paper, we aim at learning rich representations from raw EEG signals for sleep stage classification with self-supervised learning, and ask the question: \textbf{to reduce the gap caused by sampling bias, is there a better way to sample positives in SSL for sleep staging?} We answer the question by introducing a novel positive mining mechanism.

Our basic idea is to utilize prior expert knowledge to mine more semantically similar positives. It's inspired by the fact that many machine learning methods exploit domain expert knowledge to improve performance. For example, various dedicated features are extracted according to expert experience, such as power spectral density \cite{app10217639} and differential entropy \cite{duan2013differential}. The success of these methods illustrates that these handcrafted features contain semantic information of raw data. Based on above observations, we improve the conventional training regime of self-supervised learning by sampling multiple positives according to feature similarity. In detail, we retrieve top-$K$ samples having the most similar feature as positives, rather than one augmented positive or temporally neighbouring positives. 


Although mining positives based on prior knowledge alleviates sampling bias, it leads to another problem (i.e. retrieving top-$K$ samples having the most similar feature as positives may include some semantically distinct samples, which are  supposed to be negatives). This problem is unavoidable since the ground-truth labels are inaccessible,  but we further propose a weight adjustment mechanism to alleviate it. Although we do not know definitively the correctness of the selected positives, we know the relative likelihood of true positives. For a given sample, a higher similarity of the feature brings higher confidence of being a positive, vice versa. To assign higher weights to confidence positives/negatives, we utilize the property of temperature in contrastive loss. We show  by analyzing gradient in methodology section that temperature in contrastive loss affects penalty strength. By setting adaptive temperatures for each sample based on their confidence levels, high confidence samples make greater contributions and low confidence samples make a relatively smaller impact. Experiments demonstrate that the proposed adaptive temperature mechanism leads to better result. 

Key contributions of our work are summarized as follow: 
\begin{itemize}
    \item We propose a self-supervised approach, called SleepPriorCL, for sleep stage classification that utilizes prior knowledge to discover potential positives. 
    \item  By analyzing the gradient of the contrastive loss, we observe the effect of temperature on the gradient strength and propose an adaptive temperature mechanism to further improve performance.
    \item We thoroughly validate the effectiveness of the learned representations by comparing with baseline approaches. Experimental results demonstrate superior performance of our method.
\end{itemize}

\section{Related Work}
\subsubsection{Self-supervised Learning.} 
Recent success of self-supervised learning stems from the use of discriminative contrastive loss on data samples \cite{he2020momentum,chen2020simple}. Given an anchor data sample, the objective discriminates its semantic-similar positives  (e.g. obtained by data agumentations) against negative samples. Typically, negatives are sampled randomly, which may contain potential positives.
\citet{DCL} call this problem sampling bias and mitigate it by taking the viewpoint of Positive-Unlabeled learning. Some methods incorporate  multiple views of data to mine potential positives \cite{NegativeCancellation, han2020self}. However, none of the above researches explicitly propose to exploit prior knowledge for positive mining.

For time series, a line of  SSL methods sample positives from temporally neighbors  \cite{oord2018representation, franceschi2019unsupervised, tonekaboni2021unsupervised, Eldele2021TimeSeriesRL}. Specifically for physiological time series, \citet{10.1088/1741-2552/abca18} construct a binary classification task, identifying whether the selected samples are adjacent to each other. SleepDPC \cite{DBLP:conf/icassp/XiaoWYZBZW21} learn representations for sleep staging by predicting near future positives and discriminating temporal neighbors. All of the above time series SSL methods can only mine temporally neighboring positives, ignoring all  non-temporally neighboring semantic-similar positives. 
\subsubsection{Sleep Staging.} 
To achieve automatic sleep staging, a wide range of methods have been proposed \cite{aboalayon2016sleep}. Handcrafted feature based supervised methods \cite{en2014ACS, DBLP:journals/bspc/LiuSZR16} pioneered the way. Recent deep supervised neural networks \cite{Phan2019SeqSleepNetEH, GraphSleepNet, Phan2021XSleepNetMS} based on polysomnogram (PSG) achieve promising performance. However, the multi-channel PSG suffers the drawback of complicated preparation and disturbance to participants’ normal sleep, preventing it's wider usage. Thus, a branch of supervised approaches focus on single-channel EEG \cite{Seifpour2018ANA,Sors2018ACN, tiny9176741, Fu2021DeepLI}. However, supervised methods rely heavily on labels, which are laborious to obtain in medical field. Recent SSL methods demonstrate promising results when few labels are accessible \cite{DBLP:conf/icassp/XiaoWYZBZW21, 10.1088/1741-2552/abca18, Eldele2021TimeSeriesRL}.






\section{Preliminaries}
In our study, we denote the raw  EEG signal set as  $\boldsymbol{X}=\left\{x_{1}, x_{2}, \ldots, x_{L}\right\}\in \mathbb{R}^{L \times  N}$ where $L$ denotes the number of sleep epochs and  $N$ denotes the length of each sleep epoch $x_{i} \in \boldsymbol{X}(i \in\{1,2, \cdots, L\})$.
 The contrastive learning framework we used comprises the following major components.
\begin{itemize}
\item Data augmentation module $Aug(\cdot)$. For each sleep epoch $x$, we generate two augmented samples, each is denoted as $\hat{x}=Aug(x)$. In this module, EEG signals are randomly mask and randomly scaled, which provide different views of raw signals. 
\item Base encoder $f(\cdot)$, which maps a sleep epoch $x$ to a representation vector $h=f(x)$. Two augmented samples are input to encoder $f(\cdot)$, generating two representation vectors. In this paper, we use a simple 4-layer convolutional neural network as the base encoder. 
\item Projection head $g(\cdot)$, which maps $h$ to a vector $z=g(h)$. It is discarded after contrastive learning. We set it as a MLP with one hidden layer.
\item Linear classifier $c(\cdot)$, which maps $h$ to a label $\hat{y}=c(h)$. It is used to verify the validity of the learned representation under the linear evaluation protocol\cite{chen2020simple}.
\end{itemize}
The goal of this paper is to learn semantic representations from raw EEG signals by training an encoder $f(\cdot)$, with a specific application on sleep staging. The contrastive loss is conducted upon $z$. 
\section{Methodology}

\subsection{Sampling Bias in Contrastive Learning}

Following the setup of SimCLR \cite{chen2020simple}, we construct a similar framework for sleep stage classification. The loss function is formulated as:
\begin{equation}
\mathcal{L}\left(x_{i}\right)=-\log \frac{\exp \left(s_{i, p} / \tau\right)}{\exp \left(s_{i, p} / \tau\right)+\sum\limits_{n \in N(i)}\exp \left(s_{i, n} / \tau\right)}\label{singlepositiveloss}
\end{equation}
where  $s_{i,j}$ is the similarity of $z_i$ and $z_j$, measured by cosine similarity $s_{i,j}={z_i}^{\top} {z_j} /\|{z_i}\|\|{z_j}\|$. The index $i$ is called the \emph{anchor}, index $p$ is called the \emph{positive}, $N(i)$ is the set of all negatives in the mini-batch and index $n$ is called the \emph{negative}. Typically, for an anchor, the only positive is the augmented sample, and negatives are all other samples within the same mini-batch. Optimizing this loss function attract the augmented positive to anchor. However, all potential semantically similar positives within the same mini-batch are repelled, leading to sub-optimal performance. 

The above sampling bias problem can be solved if we can find all positives. Under an ideal situation, suppose all ground-truth labels are accessible, we can find all positives and attract them to the anchor. we refer to this as \emph{unbiased} contrastive learning. The loss function can be formulated as:
\begin{small}
\begin{equation}
\mathcal{L}\left(x_{i}\right)=\frac{-1}{|P(i)|} \sum_{p \in P(i)}\log \frac{\exp \left(s_{i, p} / \tau\right)}{ \exp \left(s_{i, p} / \tau\right)+\sum\limits_{n \in N(i)} \exp \left(s_{i, n} / \tau\right)}\label{multipositiveloss}
\end{equation}
\end{small}where $P(i)$ is the positive set containing all ground-truth positives of $x_i$ in the mini-batch distinct from $i$, and $|P(i)|$ is its cardinality. Note that when $|P(i)|=1$, Eq. \ref{multipositiveloss} is identical to Eq.  \ref{singlepositiveloss}. 

However, in practice, our goal is to learn meaningful representation without labels. To alleviate the sampling bias problem, is there a better unsupervised way to discover more positives?

\subsection{Incorporating Prior Expert Knowledge to Mine More Positives}
Our idea is to imitate domain experts to mine more positives. Some handcrafted features \cite{app10217639, duan2013differential} contain expert knowledge, which can be used for positive mining. In this paper, for the sleep staging task, we use a common-used feature, the signal energy of each EEG rhythm.

According to AASM rules \cite{berry2012rules}, EEG rhythms play an important role in sleep stage classification. In clinical medicine, physicians focus on four rhythms when staging a patient's sleep by EEG. These rhythms are $\delta$ rhythm (1-4Hz), $\theta$ rhythm (4-8Hz), $\alpha$ rhythm (8-13Hz) and $\beta$ rhythm (14-30Hz). Table \ref{table1} shows the major EEG rhythms of each sleep stage, 
demonstrating that different sleep stages have different EEG rhythm composition.  Therefore, we can use the energy of these EEG rhythms to discover more positives. 
\begin{table}[h]
\centering
\resizebox{.95\columnwidth}{!}{
\begin{tabular}{c|cccc}
\hline
Sleep stage & $\delta$(1-4Hz) & $\theta$(4-8Hz) & $\alpha$(8-13Hz) & $\beta$(14-30Hz) \\ \hline
W           &       &       & \checkmark     & \checkmark    \\
N1          &       & \checkmark     & \checkmark     &      \\
N2          &       & \checkmark     &       &      \\
N3          & \checkmark     &       &       &      \\
REM         &       & \checkmark     & \checkmark    &      \\ \hline
\end{tabular}
}
\caption{Major EEG rhythms of each sleep stage.}

\label{table1}
\end{table}

Following \cite{DBLP:journals/bspc/LiuSZR16}, given an epoch of EEG signal $x(n)$, method of extracting EEG rhythms energy can be described as two steps:\\
(1) Apply FFT on EEG signal $x(n)$, and get the frequency spectrum $P(\omega)$ of $x(n)$.
\begin{equation}
P(\omega)=\sum_{n=0}^{N-1} x(n) e^{-j \frac{2 \pi}{N} \omega n}, \omega=0,1, \ldots N-1
.\end{equation}
(2) Calculate signal energy of each rhythm.
\begin{equation}
\begin{aligned}
& E(\delta)=\int_{1}^{4} P(\omega)^{2} d \omega, \quad  & E(\theta)=\int_{4}^{8} P(\omega)^{2} d \omega,  \\
& E(\alpha)=\int_{8}^{13} P(\omega)^{2} d \omega,\quad   & E(\beta)=\int_{14}^{30} P(\omega)^{2} d \omega.
\end{aligned}
\end{equation}
For every epoch of EEG signal $x$, we can get a vector of energy of EEG rhythms $E=[E(\delta), (\theta), E(\alpha), E(\beta) ]$. For the remainder of this paper, we refer to $E$ as a prior feature. To some extent, the similarity of prior features represents the semantic similarity. For clarity, we define dissimilarity $d_{i, j}$ between anchor $x_i$ and sample $x_j$ as:
\begin{equation}
d_{i, j}=\log{(\left\|E_{i}-E_{j}\right\|_{2})}
\end{equation}

\begin{figure}[h]
\centering
\includegraphics[width=0.98\columnwidth]{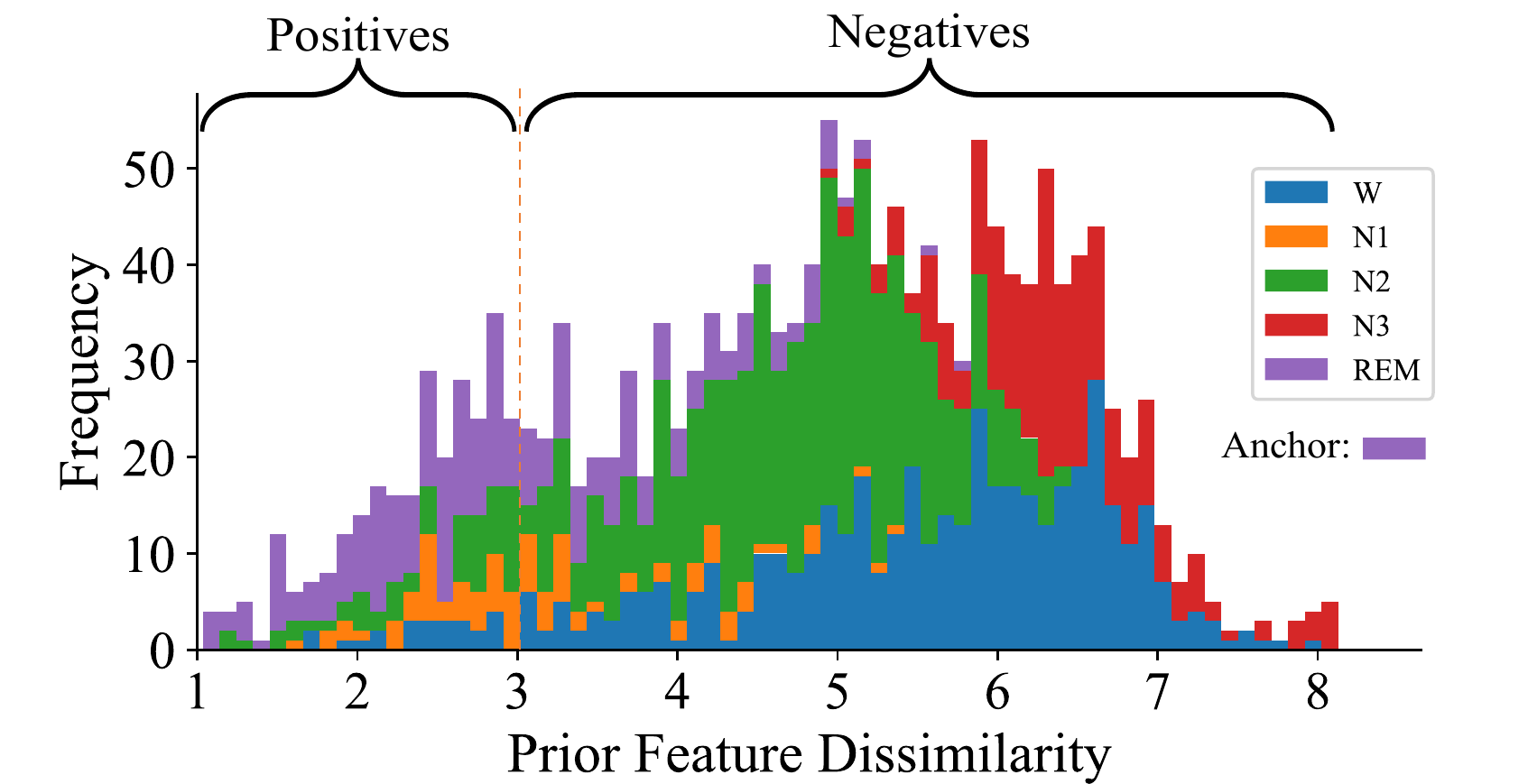} 
\caption{Histogram showing an example of mining positives. The anchor $x_i$ is a REM stage epoch, we sort samples by dissimilarity with $x_i$ and set top-$K$ as positives, the rest as negatives (i.e. $|P(i)| = K$). Most of positives are semantic-similar with the anchor (i.e. REM stage).}  
\label{anchor4}
\end{figure}

For an anchor $x_i$, discovering positives takes two steps:
(1) calculate dissimilarities $d_{i,j}$ between anchor $x_i$ and all other samples within the same mini-batch. 
(2) sort samples by dissimilarity and set top-$K$ as positives, the rest as negatives. Figure \ref{anchor4} gives an example that can help understand this process.

 However, as Figure \ref{anchor4} shows, some of the mined positives are incorrect, which could lead to a performance drop. How can we alleviate this problem? Although we do not know exactly the correctness of the mined positives, we know their relative likelihood. In detail, the smaller the dissimilarity, the higher the confidence  for  a  positive.  On  the  contrary, the greater the dissimilarity, the higher the confidence for a negative. We want high confidence samples to make greater contributions, and low confidence samples make less impact. We achieve this by introducing a mechanism that adjusts gradient penalty strength for each sample depending on their confidence level of being positive or negative.

\subsection{Contrastive Learning with Adaptive Temperature}

To adjust the strength of gradient penalty, each sample was given a customized temperature. The multi-positive contrastive loss is modified as:


\begin{small}
\begin{equation}
\mathcal{L}\left(x_{i}\right)=\frac{-1}{|P(i)|} \sum_{p \in P(i)}\log \frac{\exp \left(s_{i, p} / \tau_{p}\right)}{ \exp \left(s_{i, p} / \tau_{p}\right)+\sum\limits_{n \in N(i)} \exp \left(s_{i, n} / \tau_{n}\right)}\label{multipositive}
\end{equation}
\end{small}

Next, we  discuss the role the temperature in contrastive loss by analyzing the gradient. We  show that the temperature controls the gradient magnitude of both positives and negatives, especially hard positives/negatives (i.e., ones against
which continuing to contrast the anchor greatly benefits the encoder). Specifically, the gradients with respect to the positive similarity $s_{i,\hat{p}}$ and the negative similarity $s_{i,\hat{n}}$ are formulated as:
\begin{equation}
\frac{\partial \mathcal{L}\left(x_{i}\right)}{\partial s_{i, \hat{p}}}=\frac{-1}{\tau_{\hat{p}}|P(i)|}  \frac{\sum\limits_{n \in N(i)} \exp \left({s_{i, n}}/{\tau_{n}}\right)}{\exp \left({s_{i, \hat{p}}}/{\tau_{\hat{p}}}\right)+\sum\limits_{n \in N(i)} \exp \left({s_{i, n}}/{\tau_{n}}\right)}\label{positivegradient}
\end{equation}
\begin{small}
\begin{equation}
\frac{\partial \mathcal{L}\left(x_{i}\right)}{\partial s_{i, \hat{n}}}=\frac{1}{\tau_{\hat{n}}|P(i)|} \sum_{p \in P(i)} \frac{\exp \left(s_{i, \hat{n}} / \tau_{\hat{n}}\right)}{\exp \left({s_{i, p}}/{\tau_{p}}\right)+\sum\limits_{n \in N(i)} \exp \left({s_{i, n}}/{\tau_{n}}\right)}\label{negativegradient}
\end{equation}
\end{small}
\begin{figure}[t!]
\centering
\includegraphics[width=.95\columnwidth]{figs/gradient_pos_spe_1.pdf} 
\caption{An example of gradient w.r.t. different $s_{i, \hat{p}}$. In this example, $|P(i)|=10$, $|N(i)|=100$ and except for $\tau_{\hat{p}} \in \{0.05, 0.07, 0.1, 1.0\}$, temperature of all other samples are fixed as $\tau=0.1$.}
\label{gradient_pos_spe_1}
\end{figure}

\begin{figure}[t]
\centering
\includegraphics[width=.95\columnwidth]{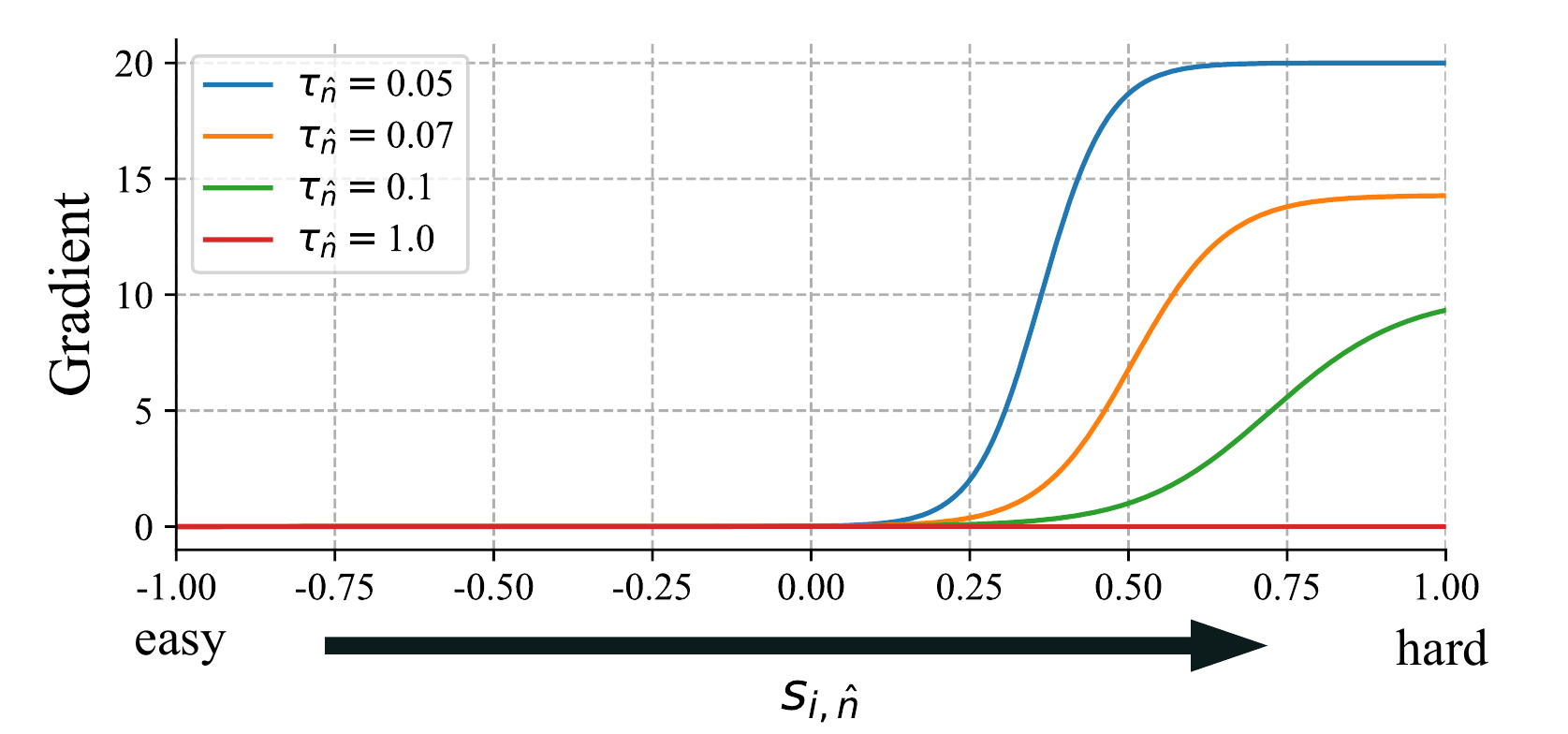} 
\caption{An example of gradient w.r.t. different $s_{i, \hat{n}}$. In this example, $|P(i)|=10$, $|N(i)|=100$ and except for $\tau_{\hat{n}} \in \{0.05, 0.07, 0.1, 1.0\}$, temperature of all other samples are fixed as $\tau=0.1$. }  
\label{gradient_neg_spe_1}
\end{figure}

Figure \ref{gradient_pos_spe_1} and Figure \ref{gradient_neg_spe_1} visualize examples of the gradient with respect to $s_{i,\hat{p}}$ and  $s_{i,\hat{n}}$ according to Eq. \ref{positivegradient} and Eq. \ref{negativegradient}.  Note that the signs of ${\partial \mathcal{L}\left(x_{i}\right)}/{\partial s_{i, \hat{p}}}$ and ${\partial \mathcal{L}\left(x_{i}\right)}/{\partial s_{i, \hat{n}}}$ are opposite, we concentrate on the magnitude of gradients. Although the graphs drawn using different parameters are not exactly the same, it does not affect our following observations:
(1) the contrastive loss is hardness-aware. The harder the sample, the greater the gradient penalty strength. Specifically, for a positive $s_{i,\hat{p}}$, if it's an easy positive (i.e. $s_{i,\hat{p}}\approx1$), the gradient magnitude is almost 0. On the contrary, if it's a hard positive (i.e. $s_{i,\hat{p}}\not\approx1$), the gradient magnitude substantial increase. A similar phenomenon can be observed 
on negatives. For a negative $s_{i,\hat{n}}$, if it's an easy negative (i.e. $s_{i,\hat{n}}\not\approx1$), the gradient is almost 0 and if it's a hard negative (i.e. $s_{i,\hat{p}}\approx1$), the gradient magnitude substantial increase.
(2) temperature has a strong impact on gradient of hard samples, but have a weak impact on gradient of easy samples. For hard positive/negative, the gradient penalty significantly increases as the temperature decreases. The lower the temperature, the greater the magnitude of gradient. On the other hand, for easy positive/negative, temperature change can not cause a significant gradient change.

\begin{figure}[h!]
    \centering
        \subfigure[Gradient w.r.t. $s_{i,\hat{p}}$]{\includegraphics[width=0.49\columnwidth]{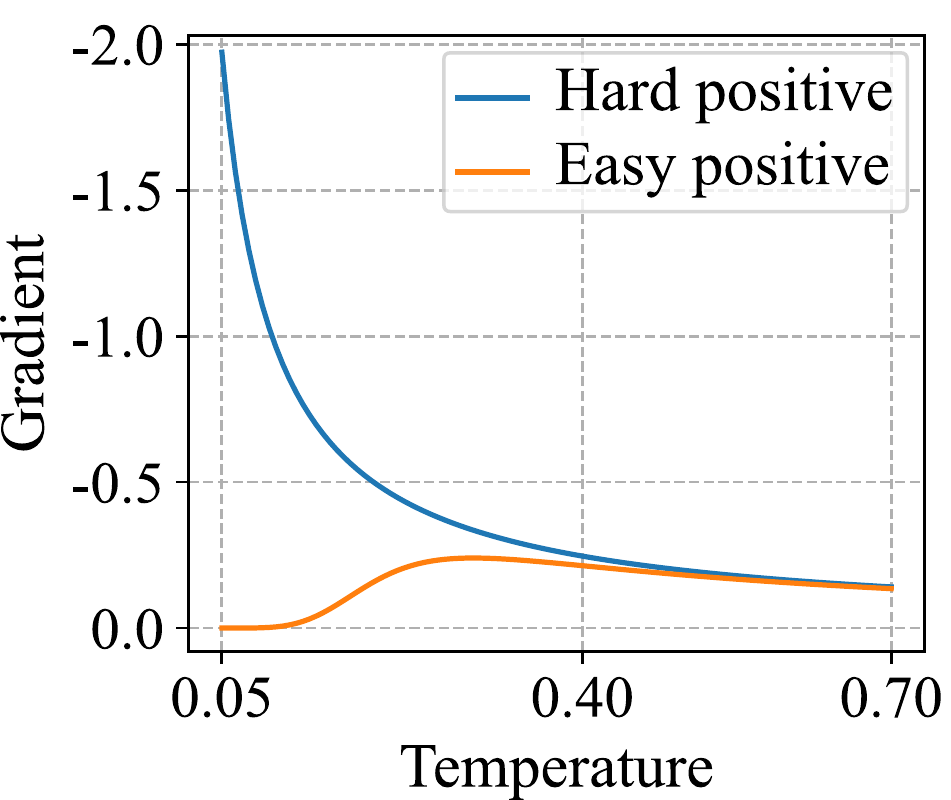}}  
        \subfigure[Gradient w.r.t. $s_{i,\hat{n}}$]{\includegraphics[width=0.49\columnwidth]{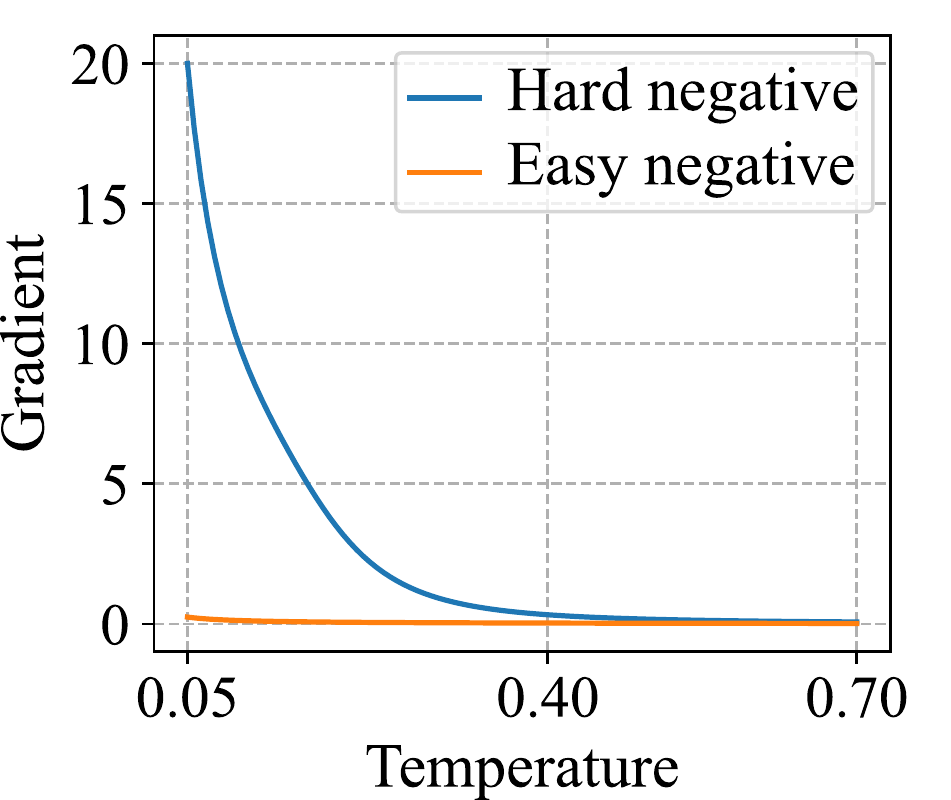}}
           \caption{Gradient of hard/easy sample.}
    \label{gradient_temperature}
\end{figure}

The above observations expose some useful properties of the optimization process for contrasting learning. As we can see in Figure \ref{gradient_temperature}, turning down the temperature can substantially increase the gradient penalty for hard samples, but has little impact on the gradient penalty for easy samples. Thus, given a positive/negative, lowering its temperature leads to two scenarios: (1) If this sample is hard, the gradient penalty will be amplified, having greater influence on the encoder training; (2) If this sample is easy, gradient penalty will be almost unchanged, having almost no impact on the encoder training. 
In conclusion, regardless of the hardness, if we want a sample to make a great contribution to the encoder training, set a low temperature for this sample. Otherwise, set a high temperature for this sample.

Based on the above conclusion, now we can adjust the weight of each sample according to their confidence levels mentioned in Section 4.2. For high confidence positives/negatives, we expect them to contribute more to the encoder training, so we give them low temperatures. On the contrary,  for low confidence positives/negatives, we expect them to make less impact, so we give them high temperatures.

\begin{figure}[t]
\centering
\includegraphics[width=0.9\columnwidth]{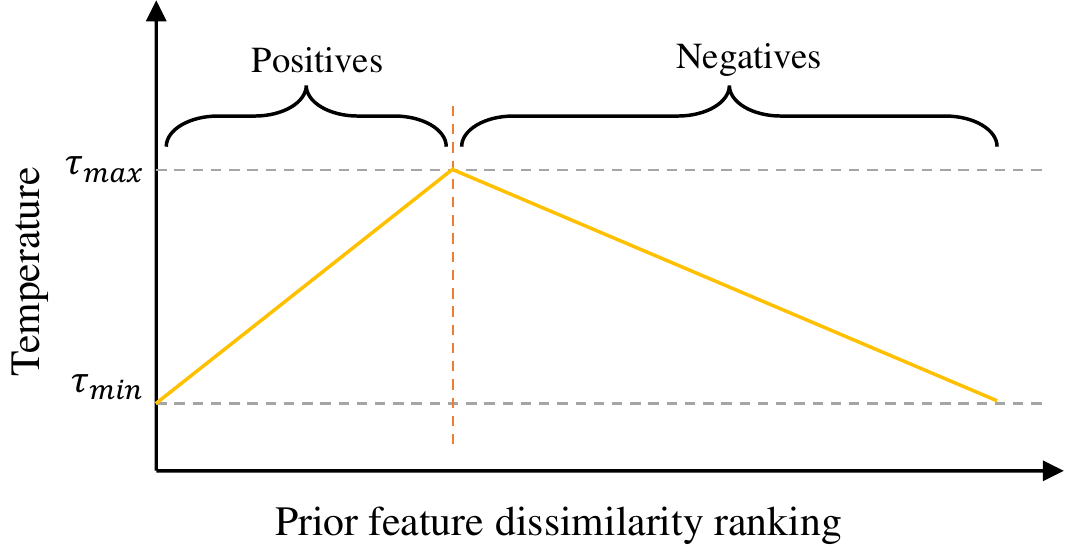}
\caption{Adaptive temperatures based on prior feature dissimilarity.}  
\label{set_temperature1}
\end{figure}

In practice, We follow these steps to mine positives and adjust the gradient penalty of each samples: (1) Sort samples by prior feature dissimilarity. (2) Set the top-$K$ samples with the smallest dissimilarity as positives, and set the rest as negatives. (3) Set the temperature of positive $x_p$ as $\tau_{p}$, and set the temperature of negative $x_n$ as $\tau_{n}$, where
\begin{equation}
\begin{aligned}
\tau_{p}=\tau_{min}+rank(x_p)\cdot\frac{(\tau_{max}-\tau_{min})}{|P|}\\
\tau_{n}=\tau_{max}-rank(x_n)\cdot\frac{(\tau_{max}-\tau_{min})}{|N|}
\end{aligned}
\end{equation}
$\tau_{min}$ and $\tau_{max}$ are the minimum and maximum temperatures we set, $|P|$ is the cardinality of positive set, $|N|$ is the cardinality of negative set, $rank(x_p)$ is the prior feature dissimilarity ranking of positive  $x_p$ in the positive set and $rank(x_n)$ is the prior feature dissimilarity ranking of negative $x_n$ in the negative set. Our proposed approach can be understood more intuitively from 
Figure \ref{set_temperature1}.

\section{Experiments}


\subsection{Datasets}
To verify the effectiveness of our method, we conducted experiments on two publicly available datasets Sleep-EDF \cite{kemp2000analysis, goldberger2000physiobank} and MASS-SS3 \cite{OReilly2014MontrealAO}. These datasets were collected using different devices with different sampling frequencies and annotated by different experts, demonstrating the generalizability of our method for the sleep stage classification.

\textbf{Sleep-EDF.} We use two subsets of Sleep-EDF, we refer to them here as Sleep-EDF39 and Sleep-EDF153. In Sleep-EDF39, sleep data were recorded from 20 healthy subjects (10 males and 10 females, 25-34 years old) with a total of 39 PSG recordings. Each subject has two nights of PSG recordings, except for the 13th subject who lost one night of sleep recording. In Sleep-EDF153, sleep data were recorded from 78 healthy subjects (37 males and 41
females,25-101 years old) with a total of 153 PSG recordings. Each subject has two nights of PSG recordings, except for the 13th, 36th and 52th subject who lost one night of sleep recording. PSG recordings in Sleep-EDF are divided into non-overlapping 30-second epochs and annotated by experts according to R\&K standard \cite{Wolpert1969AMO}. Each epoch was annotated as one of W, N1, N2,
N3, N4, REM, MOVEMENT and UNKNOWN. Fpz-Cz EEG channel is used to evaluate our method, having a sampling rate of 100Hz. As recommended in \cite{Jia2021SalientSleepNetMS}, we merge the N3 and N4 stages into a single stage N3 to use the same AASM standard as the MASS-SS3 dataset and only included 30 minutes of W periods before and after the sleep periods, as we are interested in sleep periods.

\textbf{MASS-SS3.} In MASS-SS3, sleep data were recorded from 62 healthy subjects (28 males and 34 females) with a total of 62 PSG recordings. Each subject has one night of PSG recording. PSG recordings are divided into non-overlapping 30-second epochs and annotated by experts according to AASM standard \cite{berry2012rules}. Each segments was annotated as one of W, N1, N2, N3, REM, MOVEMENT and UNKNOWN. F4-EOG (Left) channel is used to evaluate our method, having a sampling rate of 256Hz. For the above datasets, the data labeled as MOVEMENT and UNKNOWN are removed, and only the data related to sleep (W, N1, N2, N3 and REM) are retained. 


\subsection{Implementation Details}
For a fair comparison and to avoid experimental results being affected by different encoder architectures, we use a same simple encoder in SleepPriorCL and all baselines, since the objective is to compare the performance of the learning frameworks rather than network architectures. The encoder we use contains 4 one-dimensional convolutional layers, each followed by a layernorm layer and a GELU layer. Hyperparameters are $K=0.4\times batch\_size$, $\tau_{min}=0.05$ and $\tau_{max}=0.1$. For each dataset, we use single channel EEG recordings, splitting 90\% and 10\% for training and testing by subjects. To avoid the effect of randomness, each expertiment is repeated for 5 times with different 5 random seeds. During training, we use SGD optimizer with a momentum of 0.9, a learning rate of 1e-4 and a batch size of 128. The pre-training and downstream task are done for 100 and 50 epochs. We use accuracy and F1-score as metrics to evaluate the performance. All experiments are conducted using PyTorch 1.6 and GeForce RTX 2080 GPU.

\subsection{Comparison with SSL Baselines}
We compare our method with the following SSL methods under linear classifier evaluation protocol:
\begin{itemize}
\item SimCLR \cite{chen2020simple}: Conventional contrastive learing, generating only one augmented positive for each anchor.
\item DCL \cite{DCL}: Alleviate sampling bias from the viewpoint of Positive-Unlabeled learning. 
\item CPC \cite{oord2018representation}: Representation learning with contrastive predictive coding.
\item TNC \cite{tonekaboni2021unsupervised}: Unsupervised representation learning for time series with temporal neighborhood coding.
\item T-Loss \cite{franceschi2019unsupervised}: Unsupervised scalable representation learning for multivariate time series.
\item SleepDPC \cite{DBLP:conf/icassp/XiaoWYZBZW21}: Self-supervised learning for sleep staging by predicting future representations and distinguishing epochs from different epoch sequences.
\item Unbiased \cite{khosla2020supervised}: Supervised contrastive learning, pre-training encoder with all ground-truth labels.
\end{itemize}


Table \ref{linearevaluation} shows the experimental results conducted on single channel EEG of Sleep-EDF39, Sleep-EDF153 and MASS-SS3 under the linear evaluation protocol. Particularly, we train a linear classifier $c(\cdot)$ on top of a frozen self-supervised pre-trained encoder model. The results show that our SleepPriorCL outperforms all other SSL methods in both evaluation metrics, reducing the gap with unbiased method. 

The traditional contrastive leaning methods (SimCLR and DCL) treat only one augmented sample as positives, suffering the most severe sampling bias problem. On the other hand, the temporal neighbors discriminating methods (CPC, TNC, T-Loss and SleepDPC) take advantage of local smoothness of time series to discover some temporally neighboring positives, which somehow  alleviates the sampling bias problem. Therefore, the performance of temporal neighbors discriminating methods are generally better than that of traditional contrastive leaning methods.

Although temporal neighbors discriminating methods mine some temporally neighboring positives, they still ignore many other positives. Especially in physiological signals, numerous semantically similar samples are not temporally close. For example, the same sleep stage appears in the sleep recordings of different people. The propose SleepPriorCL is not constrained to temporally close positives, is also able to mine positives from the same recording but temporally distant, as well as positives from other recordings, which further alleviates the sampling bias problem. Thus, the proposed SleepPriorCL is superior to other SSL methods, 
having a minimal gap with the unbiased method.


\begin{table*}[h!]
\centering
\begin{tabular}{l|ll|ll|ll}
\toprule
                   & \multicolumn{2}{c|}{Sleep-EDF39}                              & \multicolumn{2}{c|}{Sleep-EDF153}                            & \multicolumn{2}{c}{MASS-SS3}                               \\ \midrule
Method             & \multicolumn{1}{c}{Accuracy} & \multicolumn{1}{c|}{F1-score} & \multicolumn{1}{c}{Accuracy} & \multicolumn{1}{c|}{F1-score} & \multicolumn{1}{c}{Accuracy} & \multicolumn{1}{c}{F1-score} \\ \midrule
SimCLR(Biased)             & 55.79±1.76                   & 39.44±2.12                    & 57.89±2.62                   & 29.69±1.44                    & 65.94±0.45                   & 48.96±0.87                   \\
DCL                & 52.86±4.29                   & 33.88±8.27                    & 60.93±3.45                   & 32.69±3.12                    & 62.21±0.30                   & 43.43±1.11                   \\
CPC                & 64.32±7.35                   & 48.89±9.55                    & 71.86±0.13                   & 58.54±0.39                    & 79.99±0.18                   & 68.95±0.30                   \\
TNC                & 62.27±2.17                   & 47.92±1.66                    & 64.29±1.86                   & 39.84±2.10                    & 68.43±3.90                   & 54.47±5.55                   \\
T-Loss             & 56.56±2.48                   & 38.27±3.17                    & 70.14±1.06                   & 38.98±4.01                    & 68.97±0.83                   & 52.88±1.11                   \\
SleepDPC             & 76.37±0.11                    & 62.78±0.37                     & 74.06±0.22      & 57.82±0.52                    & 79.57±0.29                    & 69.27±0.28                   \\
SleepPriorCL(Ours) & \textbf{76.44±0.83}                   & \textbf{65.56±1.32}                 & \textbf{78.11±0.43}                   & \textbf{63.60±0.80}                     & \textbf{80.40±0.26}                   & \textbf{70.60±0.75}                   \\ \midrule
Unbiased           & 79.60±0.36                   & 68.96±0.49                    & 79.17±0.11                   & 65.96±0.46                    & 83.84±0.17                   & 75.57±0.18                   \\ \bottomrule
\end{tabular}
\caption{Performance comparison under the linear classifier evaluation protocol using single channel EEG recordings.} 
\label{linearevaluation}
\end{table*}

\subsection{Comparison with Supervised Baseline}
\begin{figure}[h]
\centering
\includegraphics[width=0.9\columnwidth]{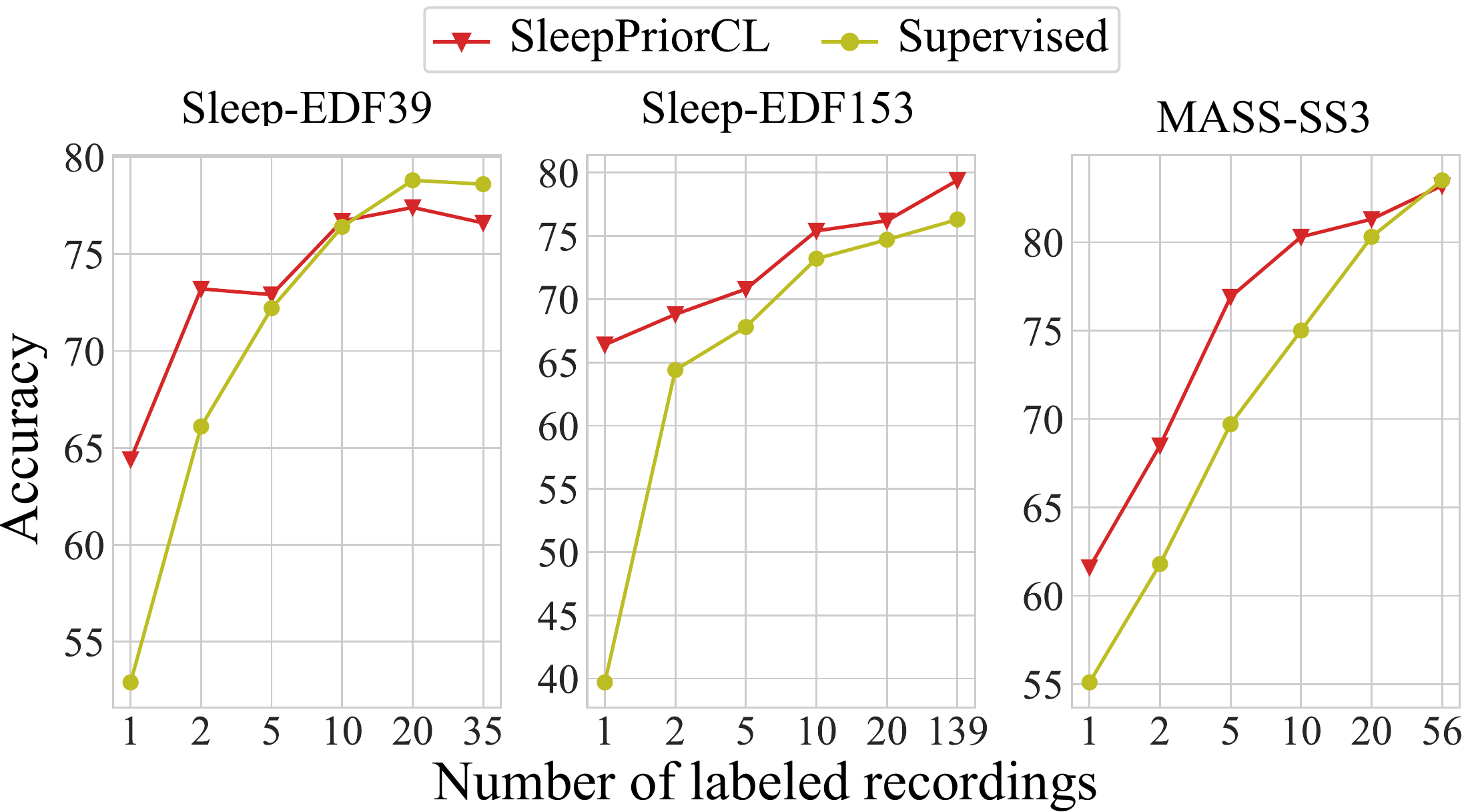}
\caption{Comparison between supervised method and fine-tuning our pre-trained encoder with different number of labeled single channel EEG recordings. }  
\label{semi}
\end{figure}

We also compare the performance between our method and supervised method  with a fraction of labeled data using the same simple encoder structure. We train our pre-trained encoder (SleepPriorCL) and a randomly initialized encoder (Supervised)  with randomly selected 1, 2, 5, 10, 20 and all of the labeled training sleep recordings. 
Figure \ref{semi} shows the comparison of accuracy between SleepPriorCL and Supervised. 

We observe that when rare labeled single channel EEG recordings are available, the encoder pre-trained with our method is significantly superior to the supervised model.
Such a result makes sense in the medical scenario since that mass of unlabeled data is usually available, and improving sleep staging with seldom labeled sleep recordings can substantially free up the labor force.
\subsection{Analysis}

\subsubsection {Ablation Study} To further investigate the effectiveness of each module in our method, we design the following experiments conducted on Sleep-EDF153:
 
 \begin{itemize}
 \item Basic. Contrastive learning without prior knowledge. That is, the only positive is the augmented sample.
\item Feature. Supervised sleep staging using KNN based on prior feature. 
    
\item Basic+Feature. Contrastive learning that mine top-$K$ positives with prior knowledge, but without the adaptive temperature.
\item Basic+Featur+Adaptive (Ours). Contrastive learning that mine top-$K$ positives with prior knowledge and equipped with the adaptive temperature mechanism.
\end{itemize}

\begin{figure}[h]
\centering
\includegraphics[width=0.9\columnwidth]{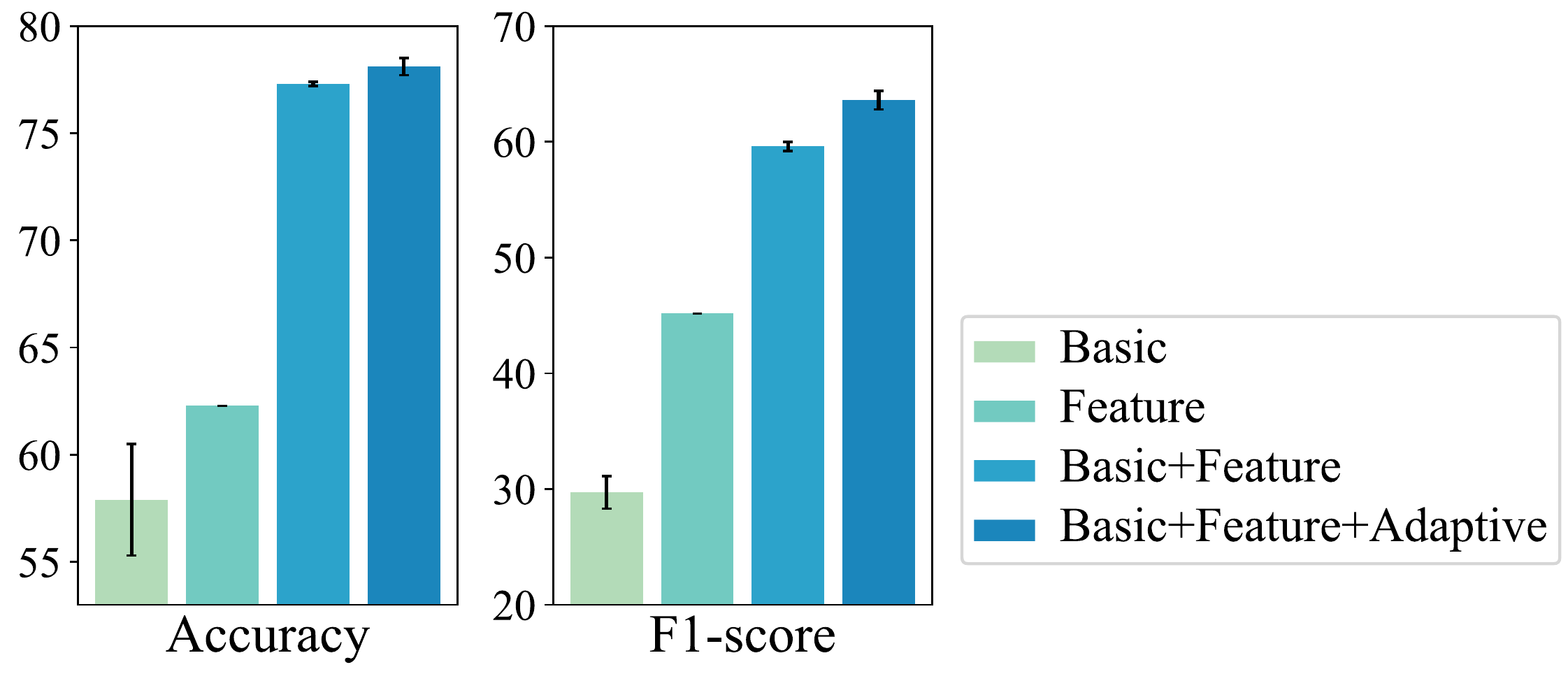}
\caption{Result of ablation study. Contrastive learing methods are evaluated under the linear classifier evaluation protocol.}  
\label{ablation}
\end{figure}

Figure \ref{ablation} demonstrates that although neither the basic contrastive learning method nor the feature-based KNN performs well, incorporating them can significantly improve performance. In other words, prior knowledge-based feature helps to discover more semantically similar positives, which alleviate the sampling biased problem. Moreover, the proposed adaptive temperature mechanism further improves the performance.
\subsubsection{Sensitivities w.r.t Hyperparameters}
 We perform  sensitivity analysis on Sleep-EDF153 to study three hyperparameters namely, the number of selected positives $K$ while retrieving the top-$K$ samples as positives, besides $\tau_{min}$ and $\tau_{max}$ in Eq. \ref{set_temperature1}.
 
\begin{figure}[h!]
    \centering
        \subfigure[$K=R\times batch\_size$]{\includegraphics[width=0.45\columnwidth]{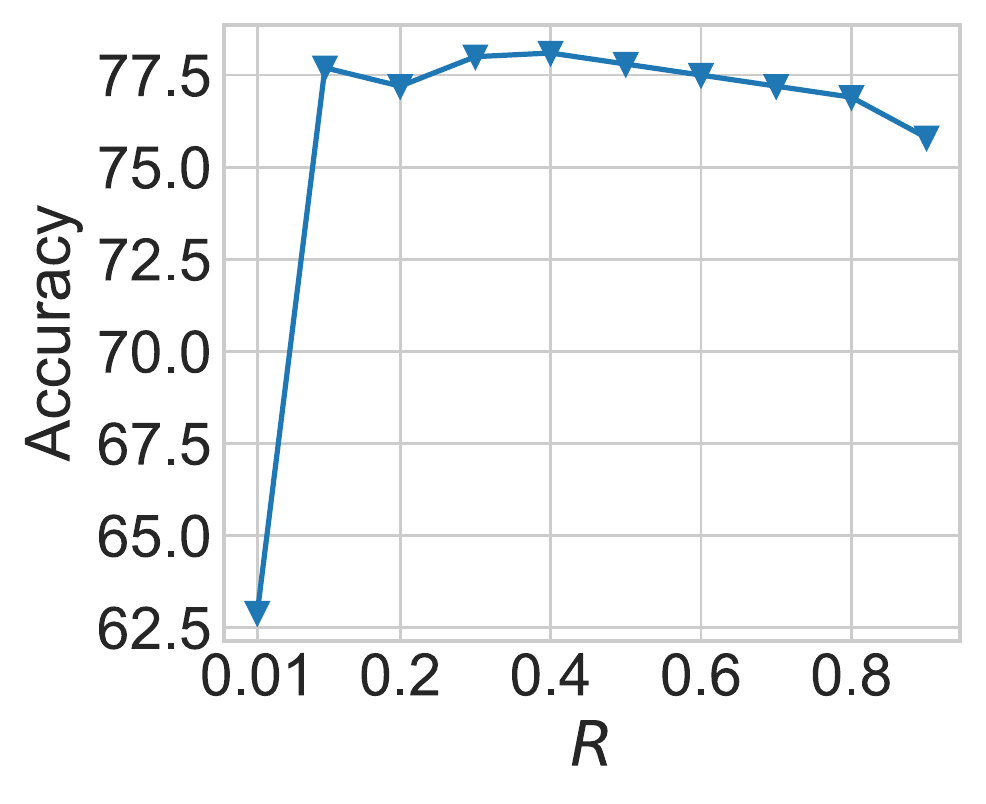}\label{sensitivityK}}  
        \subfigure[$\tau_{min} $ and $\tau_{max}$]{\includegraphics[width=0.45\columnwidth]{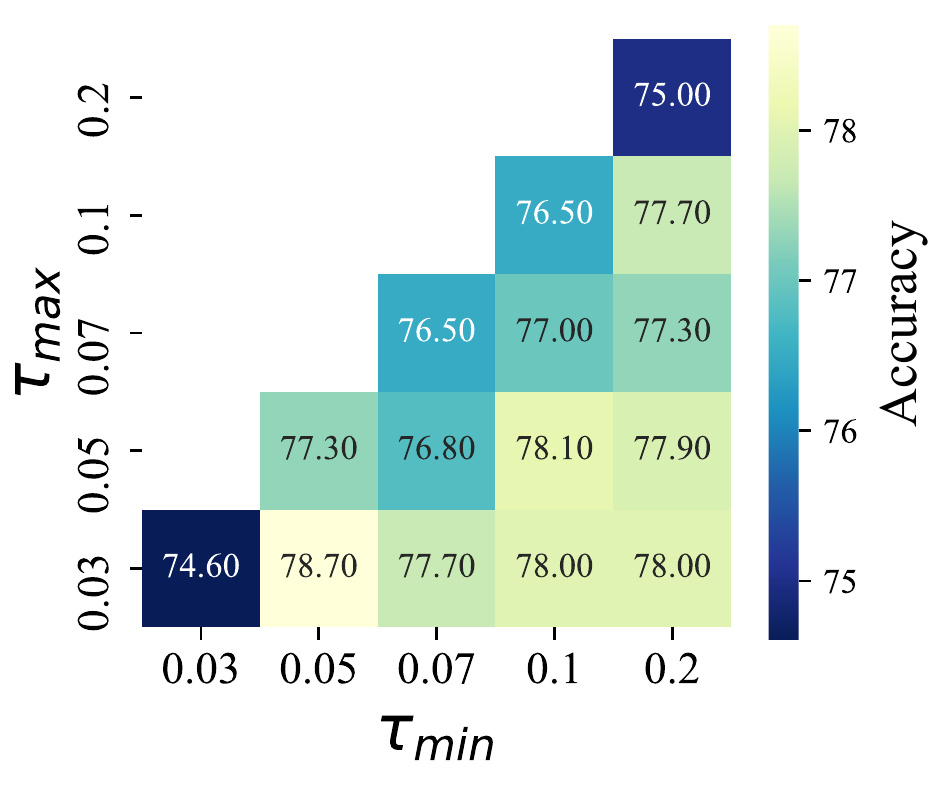}\label{sensitivitytau}}
          \caption{Sensitivities w.r.t. hyperparameters experiments on Sleep-EDF153. (a) shows the effect of $K$, where $K=R\times batch\_size$. (b) shows the  effect of $\tau_{min} $ and $\tau_{max}$.}
    \label{hyperparamenter}
\end{figure}

Figure \ref{sensitivityK} shows the effect of $K$ in the top-$K$ retrieving, where $K=R\times batch\_size$. Clearly,
when $K$ is too low ($R=0.01$), too few positives are mined, resulting in a weak ability to learn meaningful representations. A suitably $K$ ($0.1 \leq R \leq 0.4$) improves the performance, but a larger $K$ can harm the performance as it includes too many false positives. Figure \ref{sensitivitytau} shows the accuracy results for different combinations of $\tau_{min}$ and $\tau_{max}$. We can observe: 1) adaptive temperatures (results not on the sub-diagonal) generally outperform the fixed temperatures (results on the sub-diagonal); 2) under the adaptive temperature mechanism, it's not very sensitive to $\tau_{min}$ and $\tau_{max}$.

\section{Conclusion}
In this paper, we propose a novel contrastive representation learning method for sleep staging. The main novelties of the proposed method are to exploit prior domain knowledge for mining positives and adjusting each sample's gradient penalty strength. Experimental results demonstrate that our method outperforms baselines, having a promising performance using few labeled single-channel EEG recordings.


\bibliography{aaai22}

\end{document}